\documentclass[
reprint,amsmath,amssymb,aps,pra,floatfix,]{revtex4-1}
\usepackage{graphicx}	
\usepackage{dcolumn}	
\usepackage{bm}            
\usepackage{color}

\begin{document}
\title{
Precision
measurement of the 
$n$$=$$2$ 
triplet 
P
$J$$=$$1$-to-$J$$=$$0$ 
fine structure of atomic helium using 
frequency-offset 
separated oscillatory fields
}
\author{F. Heydarizadmotlagh}
\author{T.D.G. Skinner}
\author{K. Kato}
\author{M.C. George}
\author{E.A. Hessels} 
\email{hessels@yorku.ca}
\affiliation{Department of Physics and Astronomy, York University, Toronto,
Ontario M3J 1P3, Canada}

\date{\today} 

\begin{abstract}

Increasing accuracy of the
theory and experiment 
of the 
$n$$=$$2$ 
$^3$P 
fine structure of 
helium has allowed for 
increasingly-precise 
tests of 
quantum electrodynamics
(QED), 
determinations of the 
fine-structure 
constant
$\alpha$,
and limitations on 
possible
beyond-the-Standard-Model
physics.
Here we 
present a
2~part-per-billion 
(ppb)
measurement of the 
$J$$=$$1$-to-$J$$=$$0$ 
interval. 
The measurement is 
performed using 
frequency-offset
separated-oscillatory-fields.
Our result of 
$29\,616\,955\,018(60)$~Hz
represents a 
landmark for 
helium 
fine-structure
measurements,
and, 
for the first time,
will allow for 
a 
1-ppb
determination of the 
fine-structure
constant
when 
QED
theory for the interval 
is improved.

\begin{description}
\item[PACS numbers]
\verb+\pacs{32.70.Jz,32.80.-t}+
\end{description}

\end{abstract}

\pacs{Valid PACS appear here}
\maketitle

In 1964, 
Schwartz 
suggested 
\cite{PR.134.A1181} 
that a 
part-per-million 
determination
of the 
fine-structure 
constant 
$\alpha$ 
might be possible using the 
2$^3$P 
fine structure of atomic helium if 
advances were made to both 
theory and experiment. 
In the intervening decades, 
experimental measurements
have shown continual improvements
\cite{PR.169.55,
PRL.26.1613,
PRA.24.264,
PRA.24.279,
PRL.72.1802,
IEEETransInstrMeas.44.518,
PRL.84.4321,
PRL.105.123001,
OpticsComm.125.231,
PRL.82.1112,
PRL.92.023001,
PRL.97.139903,
CanJPhys.83.301,
PRL.95.203001,
PRA.58.R8,
PRL.84.3274,
PRL.87.173002,
PRA.79.060503,
PRA.91.030502,
zheng2017laser,
Kato2018ultahighprecision}.
The evaluation 
of new systematic effects
\cite{PRA.86.012510,
PRA.86.040501,
PRA.89.043403,
marsman2015quantum,
marsman2015effect}
has improved the agreement
between these measurements.
Quantum-electrodynamic 
(QED)
theory
\cite{PR.137.A1672,
PR.140.A1498,
PRA.5.2027,
PRA.6.865,PRA.7.479,
AnnPhysNY.82.89,
PRL.29.12,
PRA.18.867,
PRL.74.4791,
PRA.54.1252,
PRA.53.3896,
PRL.77.1715,
CanJPhys.80.1195,
JPhysB.33.5297,
JPhysB.36.803,
JPhysB.43.074015,
JPhysB.32.137,
PRL.97.013002,
PRA.79.062516,
PRA.80.019902,
PRL.104.070403,
CanJPhys.89.95,
JPhys.264.012007,
pei2015precision}
for these intervals
has also advanced greatly.

We present here a
measurement 
of the 
$2^3$P$_1\!\!\to\!2^3$P$_0$
interval 
that
has an uncertainty of only 
2
parts per billion
(ppb).
This work  
is the experimental contribution 
towards using the
2$^3$P 
fine structure for
tests of physics and fundamental constants 
at the 
1-ppb 
level.
An advance of
QED
theory to this same level of accuracy,
requires 
extending 
the  
calculation of 
Pachucki 
and 
Yerokhin
\cite{PRL.104.070403}
to one order higher in 
$\alpha$. 

\begin{figure}[b!]
\includegraphics[width=2.3 in]
{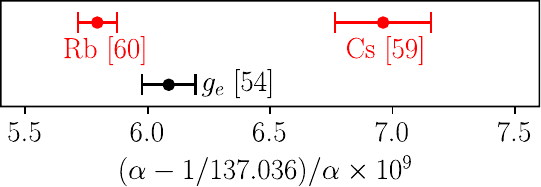}
\caption{\label{fig:alpha}
Current 
determinations of $\alpha$
obtained from atomic recoil 
in 
Cs
and 
Rb
and from 
the electron 
magnetic dipole moment
($g_e$), 
which span a range
of over 
1~ppb.
}
\end{figure}

The payoff from a
ppb-level 
comparison between 
experiment and 
QED
calculations 
will be large. 
It
will provide the most accurate test
to date of 
QED
in a 
multielectron
system
\cite{pachucki2017testing}.
This
comparison will allow the 
2$^3$P
fine structure to be used for
a direct 
test 
(at 100 times the current accuracy)
for 
beyond-the-Standard-Model
physics
\cite{pachucki2017testing}, 
such as exotic 
spin-dependent 
interactions between electrons
\cite{ficek2017constraints}.
Finally, 
the combination of 
2-ppb
theory and experiment
will allow for a determination of 
$\alpha$ 
at a level of 
1~ppb
--
one thousand times more 
accurate than the proposal of 
Schwartz 
\cite{PR.134.A1181}. 
At this precision,
the helium 
fine-structure
measurement
would provide an important 
contribution to determining
$\alpha$, 
as
the
best determinations of 
$\alpha$,
based on the electron magnetic moment
($g_e$)
\cite{
PhysRevLett.130.071801,
aoyama2012tenth,
laporta2017high,
aoyama2018revised,
PhysRevD.100.096004}
and atomic recoil
\cite{parker2018measurement,
bouchendira2011new},
currently show 
discrepancies of 
more than 1~ppb
(see
Fig~\ref{fig:alpha}).
Comparing values of 
$\alpha$
determined from various systems 
allows for tests of
beyond-the-Standard-Model
physics in each of the systems
\cite{bouchendira2011new,
PhysRevLett.130.071801}.
In particular, 
the 
$g_e$
measurement,
given another determination of 
$\alpha$,
becomes  
the most precise test 
of 
QED,
and tests
for possible substructure of the electron
\cite{bouchendira2011new,
gabrielse2014precise}
and
the possible presence of dark photons
\cite{bouchendira2011new,
hanneke2008new,
kahn2017light},
as well as putting limits on possible 
dark axial vector 
bosons
\cite{bouchendira2011new,
hanneke2008new}.
The recoil measurements,
along with another 
$\alpha$
determination,
could be used for
a mass standard 
\cite{lan2013clock}.
 
The current work is the 
most precise use to date
of the  
frequency-offset
separated-oscillatory-fields 
(FOSOF)
technique
\cite{vutha2015frequency},
which is a modification of the 
Ramsey 
method 
\cite{ramsey1949new}
of 
separated oscillatory fields
(SOF).
To implement 
FOSOF, 
the frequencies 
of the separated fields 
are slightly offset from each other, 
so that the relative phase of the 
two fields varies continuously with time.

\begin{figure*}[t!]
\includegraphics[width=160mm]
{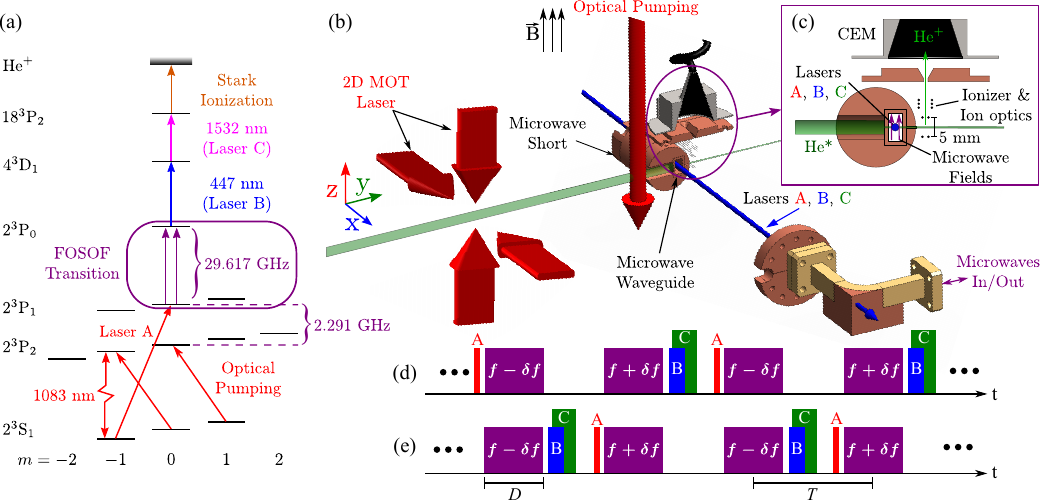}
\caption{\label{fig:ExptSetup} 
The experimental setup for the measurement.
An
energy-level
diagram 
(a) 
shows the 
29.617-GHz 
interval being measured
and the laser
transitions used
for the 
2D 
magneto-optical 
trap,
for optical pumping,
and for the three laser pulses 
(A, B, C).
The experimental setup 
(b), 
along with an expanded view of 
the region where the measurement
takes place 
(c),
shows the laser and microwave 
interactions and ionization detection.
The timing diagrams, 
(d) 
and
(e),
show the laser
and 
FOSOF 
microwave pulses
for the two timing sequences
used. 
}
\end{figure*}

Many aspects of our measurement apparatus
(see Fig.~\ref{fig:ExptSetup})
are similar to that used
\cite{Kato2018ultahighprecision}
for our measurement of the 
$2^3$P$_2\!\!\to\!2^3$P$_1$ 
interval.
Therefore we give only a brief 
description 
here.
A beam of 
metastable
2$^3$S
helium
atoms is created in a 
liquid-nitrogen-cooled
DC 
discharge source
and is 
intensified  
by a 
two-dimensional
magneto-optical 
trap.
The atoms are 
optically pumped 
(see
Fig.~\ref{fig:ExptSetup})
into the 
2$^3$S($m$=$-$1) 
state before passing through a 
0.5-mm-diameter 
hole in 
a rectangular microwave 
waveguide
(WR-28, with inner dimensions of 
7.112
by
3.556~mm).
The measurement takes place inside this 
waveguide.
The 
2$^3$S($m$=$-$1)
atoms are excited by a 
pulse of 
1083-nm
laser light 
(A 
in 
Fig.~\ref{fig:ExptSetup})
to the
2$^3$P$_1$($m$=0)
state.
The 
$2^3$P$_1(m$=0)$\to$2$^3$P$_0(m$=0)
transition is then driven
with 
29.6-GHz
microwaves.
The resulting
2$^3$P$_0$
atoms are detected 
via
excitation to
4$^3$D$_1$
using a 
pulse of
447-nm
laser light
(B
in 
Fig.~\ref{fig:ExptSetup}),
and then to
18$^3$P$_2$
using a 
pulse of 
1532-nm light
(C
in 
Fig.~\ref{fig:ExptSetup}).
The 
18$^3$P$_2$
atoms are 
Stark-ionized
by electric fields 
(see 
Fig.~\ref{fig:ExptSetup}(c)),
with the resulting ions 
being
focused 
through a 
slit into a 
channel electron multiplier.

Diode lasers are used
to produce all three wavelengths,
with 
fiber
amplifiers being used for  
1083 nm
and 
1532 nm.
Laser pulses are created
using double passes 
through 
acousto-optic 
modulators. 
The transition being measured 
is driven with two pulses of microwaves,
each of duration 
$D$, 
and separated in time by 
$T$,
as shown in 
Fig.~\ref{fig:ExptSetup}(d). 
The pulses are created by fast switching of
microwaves output from two 
precision generators,
with their internal clocks locked to 
each other and referenced to both 
Rb 
and 
GPS
clocks.
The 
amplified
microwaves
(of power $P$$\leq$2~W)
enter one end of the 
waveguide
and reflect off of a short 
that is situated 
one half wavelength
from the 
0.5-mm 
hole. 
This forms a standing wave
that has an 
antinode
of 
magnetic field
and a node
of electric field
along the 
center 
line of the 
atomic beam.

\begin{figure}
\includegraphics[width=2.8in]
{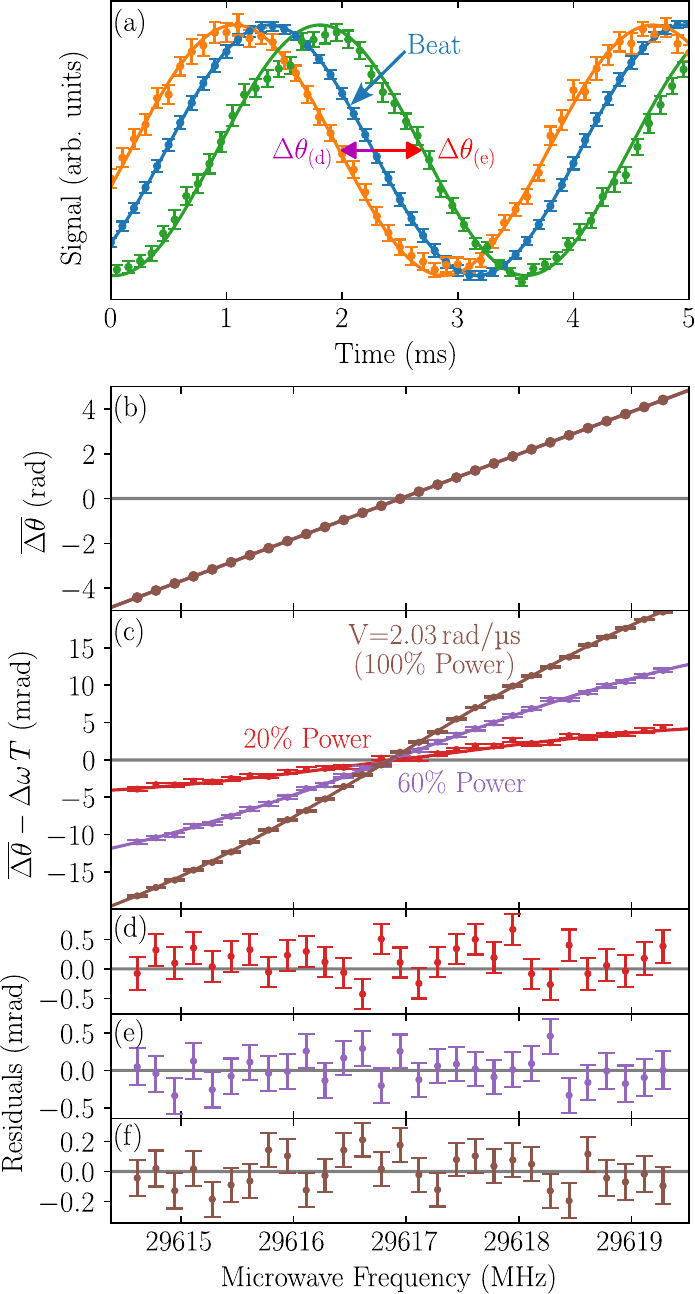}
\caption{\label{fig:LineShape} 
The FOSOF lineshape.
The sinusoidal atomic signals for 
the configurations of 
Fig.~\ref{fig:ExptSetup}(d)
and
(e)
are shifted by 
$\Delta\theta_{\rm (d)}$
and
$\Delta\theta_{\rm (e)}$ 
relative to a 
microwave beat signal, 
as shown in 
(a).
$\overline{\Delta\theta}$ 
=($\Delta\theta_{\rm (d)})$$-$$
\Delta\theta_{\rm (e)}$)/2
is shown in 
(b).
A 
300 
times expanded scale in 
(c),
where  
$\Delta\omega\,T$
is subtracted,
resolves the lineshapes
for different powers.
The fits in 
(c) 
use
Eq.~(\ref{eq:LineShape}),
and the residuals from the fits 
are shown in 
(d), 
(e),
and 
(f).
}
\end{figure}

The microwave frequencies
of the pulses are offset by
$\pm\delta\! f$,
with pulses alternating between 
$f$$+$$\delta\! f$
and 
$f$$-$$\delta\! f$.
The offset frequency 
$2\,\delta\! f$
causes the relative phases of the two
pulses to vary continuously in time.
As a result,
the atomic signal 
(see 
Fig.~\ref{fig:LineShape}(a))
varies sinusoidally in time, 
cycling between destructive and 
constructive interference.
The phase difference 
$\Delta\theta$
between this 
signal and a beat signal
obtained by combining the microwaves
at the two frequencies 
is shown in 
Fig.~\ref{fig:LineShape}(a).
We take data with 
two different timing sequences:
Fig.~\ref{fig:ExptSetup}(d),
in 
which the 
$f$$+$$\delta\! f$ 
pulse occurs before the 
$f$$-$$\delta\! f$ 
pulse,
and 
Fig.~\ref{fig:ExptSetup}(e),
in which this ordering is reversed.
To switch from  
(d)
to
(e),
only the timing of the laser pulses
is shifted -- 
the microwave pulses remain unchanged.
Fig.~\ref{fig:LineShape}(a)
shows that the
sign of the
phase shift 
$\Delta\theta$
is opposite for the two cases,
and
$\overline{\Delta\theta}
$=($\Delta\theta_{\rm (d)})$$-$$
\Delta\theta_{\rm (e)}$)/2
cancels 
unintended phase shifts 
in both the atomic
and 
beat
signals
\cite{vutha2015frequency}.

For the simple case of a two-level
system with two ideal pulses of 
duration 
$D$ 
and separation 
$T$,
the 
FOSOF 
lineshape 
is
\begin{equation}
\label{eq:LineShape}
\overline{\Delta \theta} (f)
=
\Delta\omega(T\!-\!D)
\!+\!2\tan\!^{-1}\!
\big[
\frac
{\Delta \omega 
\tan(\sqrt{4 V^2\!+\!\Delta\omega^2}D/2)}
{\sqrt{4 V^2\!+\!\Delta\omega^2}}
\big],
\end{equation}
where 
$V$ 
is the
magnetic-dipole
matrix element driving the transition,
and 
$\Delta \omega$/2$\pi$$=$$f$$-$$f_0$ 
is the separation between the applied microwave
frequency and the atomic resonant frequency. 
This line shape is 
antisymmetric
with respect to
$\Delta \omega$
(with 
$\overline{\Delta \theta}$$=$0
at 
$\Delta \omega$$=$0), 
and reduces to the very simple
proportional relation
$\Delta \omega \, T$
for small
$V$.
The observed lineshape is shown in 
Fig.~\ref{fig:LineShape}(b) for 
the case of 
$T$$=$300~ns
and
$D$$=$100~ns
for three different values of 
$V$
(i.e.,
different powers 
$P$).
The 
lineshapes are 
very nearly described by the  
$\Delta \omega \, T$
linear 
expression.
On a
300-times 
expanded scale in 
Fig.~\ref{fig:LineShape}(c),
where the 
$\Delta \omega \, T$
straight line is 
subtracted,
one can see that the data is described 
well by the lineshape of 
Eq.~(\ref{eq:LineShape}).


A fit to the 
data taken at 
100\% power
(corresponding to 
$V$$=$$2.03$~rad/$\rm{\mu}$s) 
gives 
an
$f_0$
determination
with an uncertainty of only 
13 Hz.
The residuals from this 
fit,
shown
in 
Fig.~\ref{fig:LineShape}(f),
are 
$100\,000$
times smaller than
the range shown in 
Fig.~\ref{fig:LineShape}(b),
indicating that the lineshape
is understood at 
this level.
The individual points 
in this fit are averages
of 
100~minutes of data
(taken at various times
that span six
months of data collection). 
The excellent 
signal-to-noise
ratio
is despite the fact that 
the measurement
sequence takes 
450~ns
(much longer than the
98-ns 
2$^3$P
lifetime),
allowing only 
$e^{-(450\, {\rm ns})/(98\, {\rm ns})}$=1.0\%
of the 
2$^3$P
atoms to contribute to the signal. 

\begin{figure*}[t!]
\includegraphics[width=6.5in]
{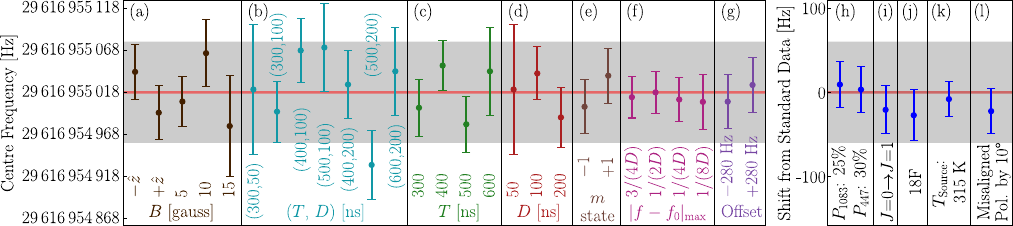}
\caption{\label{fig:Parameters} 
A summary
of the average obtained 
center
value for the various values 
of the experimental parameters. 
Details are explained in the text. 
All points use 
the 
frequency range 
$|f-f_0|$$<$$1/(2D)$, 
unless otherwise specified.
Parts
(h)
through
(l)
show shifts of  
linecenters
(for
$T$=300~ns,
$D$=100~ns,
and
100\% power)
obtained with
nonstandard parameter values 
(compared to similar
linecenters with standard
parameter values).
}
\end{figure*}

Our experiment is performed within
a magnetic field 
$\vec{B}$
of 
5~gauss, 
which 
is applied by 
20-cm-radius
Helmholtz coils,
with 
geomagnetic and other local fields
canceled 
by six larger coils.
This 
cancellation 
is calibrated by 
comparing the 
magnitude of the 
Zeeman 
shifts of our 
2$^3$P 
intervals when positive and negative 
fields are applied in each of the three
directions.
The largest systematic correction
in our measurement is a 
second-order
Zeeman 
shift
of
197.74~Hz/gauss$^2$.
The quadratic shift rate is 
precisely calculated
\cite{yan1994high},
and has been directly
tested by other measurements
\cite{lewis1970experiments,
PRL.95.203001}.
We also use
larger 
$B$
to directly show that
we understand the magnetic shifts
at a level of 
$<$0.1\%,
and we include a 
0.1\% uncertainty 
to all 
Zeeman 
corrections.
Fig~\ref{fig:Parameters}(a)
shows that measurements
taken with
$\vec{B}$ in the 
$+\hat{z}$
and
$-\hat{z}$
directions agree,
and that those with  
$|\vec{B}|$$=$5~gauss
agree with those taken for 
$|\vec{B}|$$=$10~gauss
and
$|\vec{B}|$$=$15~gauss
(which have 
four 
and
nine 
times 
larger
Zeeman 
shifts,
respectively).

\begin{figure}[t!]
\includegraphics[width=2.8in]
{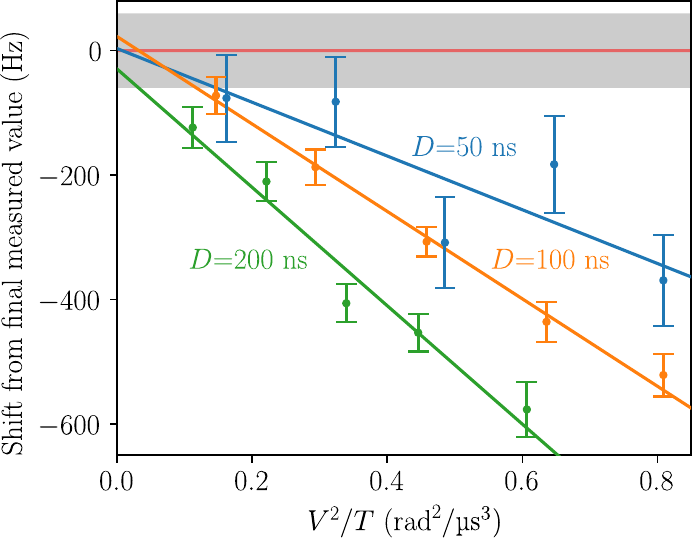}
\caption{\label{fig:LinearShifts} 
The extrapolation of the averaged
$D$$=$200~ns,
$D$$=$100~ns
and 
$D$$=$50~ns
FOSOF fit centers to 
$P$$=$0,
where the 
center 
is unaffected by 
imperfections in the pulses.
The extrapolated 
centers,
along with their uncertainties,
are shown in 
Fig.~\ref{fig:Parameters}(d).
}
\end{figure}

Eq.~(\ref{eq:LineShape}) assumes 
perfect microwave pulses, 
including sudden 
turn-on
and 
turn-off,
no chirp in the 
phase due to the 
microwave switches
or amplifier,
and no changes in intensity
or phase profiles as 
a function of 
$f$.
Imperfections in the pulses
cause the second largest
systematic effect in our measurement.
Extensive 
modeling
shows that 
all forms of distortion 
give shifts that 
are proportional to
$P$
(that is,  
shifts that extrapolate
exactly to zero in the  
zero-power
limit).
Therefore,
we linearly extrapolate 
our measured 
FOSOF 
centers
to 
$P$$=$0,
as shown
in 
Fig~\ref{fig:LinearShifts}.
These extrapolations also 
account for very small 
AC
Zeeman
and 
Stark 
shifts
caused by the 
microwaves.
Both 
modeling 
and 
measurements
show that the slopes
for these extrapolations
are nearly proportional 
to 
$D/T$.
Measurements are repeated  
for combinations of
$T$
and
$D$
to confirm that all sets of parameters 
extrapolate to a single 
intercept,
as shown in 
Fig.~\ref{fig:Parameters}(b),
(c)
and 
(d).
Both in our previous 
measurement
\cite{Kato2018ultahighprecision} 
and in the current work,
some data are also taken with
different levels of imperfections 
(and therefore slopes that differ
by a factor of approximately three),
and these still resulted in consistent intercepts
within the 
accuracy of the tests.

The Doppler 
shift is small 
since our atoms have 
an average speed of 
1100~m/s, 
as measured by their 
time-of-flight
delay 
(relative 
to the ultraviolet 
emission from the 
dc-discharge 
source)  
after passing through a 
mechanical chopper.
The average speed for atoms 
contributing to the 
FOSOF
signal is 
smaller
($\beta=v/c<3\times10^{-6}$), 
as the slower atoms are more
effectively laser excited 
and have a larger probability
of undergoing the full pulse 
sequence of 
Fig.~\ref{fig:ExptSetup}(d, e).
The
microwaves travel in a 
direction that is perpendicular
(to within 
5~mrad)
to the velocity of the atoms.
The 
Doppler 
shift is further reduced
by reflecting the 
microwaves  
from a short
and having them
intersect the atomic beam
a second time.
Ohmic
heating 
by the microwaves of the
gold-plated 
waveguide
is calculated to lead to a
difference of 
only 
0.4\%
for power of the 
reflected microwaves.
The net 
Doppler shift,
after taking these
factors into account,
is zero with an 
uncertainty of 
$\pm$2~Hz.

The polarization for the 
optical-pumping 
step is reversed for half of
the measurements, 
leading to a starting population
in the $2^3$S$(m$$=$$+1)$
state.
Fig.~\ref{fig:Parameters}(e)
shows that consistent results are 
obtained with 
$m$$=$$\pm1$.
To test for possible 
FOSOF 
lineshape 
effects, 
the data are refit with only the 
central frequencies  
($|f$$-$$f_0|$$<$$1/(2D)$,
$1/(4D)$,
or
$1/(8D)$)
included in the fit.
Consistent results are found, 
as shown in 
Fig.~\ref{fig:Parameters}(f).
Fig.~\ref{fig:Parameters}(g)
shows that the result does
not depend on the sign of the 
offset frequency.

To test for light shifts due to unintended 
temporal overlap of the laser and microwave
pulses, 
data are taken at 
lower powers for the 
447-nm 
and
1083-nm 
lasers.
As shown in 
Fig.~\ref{fig:Parameters}(h),
no shifts are seen. 
The fact that the same result 
is obtained when driving 
the 
$J$$=$$1$-to-$J$$=$$0$
and
$J$$=$$0$-to-$J$$=$$1$
(Fig.~\ref{fig:Parameters}(i))
and when using a different
Rydberg state
(the 
18$^3$F state,
Fig.~\ref{fig:Parameters}(j)),
shows that unintended atomic
processes are not affecting the 
measurement.
A warmer source temperature
(315~K,
c.f. 
130~K)
also reveals no inconsistency,
as seen in 
Fig.~\ref{fig:Parameters}(k).
Finally,
intentionally misaligning the 
linear polarization of 
laser~A 
of
Fig.~\ref{fig:ExptSetup}
by
10$^\circ$
away from the applied
$\vec{B}$ field
leads to no change in the center
(Fig.~\ref{fig:Parameters}(l)).
This latter test (along with the test
of 
Fig.~\ref{fig:Parameters}(i)
and the fact that the result 
is independent of magnetic field)
shows that transitions from 
2$^3$P$_1$($m$$=$$\pm 1$) 
to 
2$^3$P$_0$
(that would be possible if 
the microwave polarization 
were not perfect)
do not have a significant
effect on the measurement.

When all of the data used for
this measurement is averaged,
the statistical uncertainty is 
$<$5~Hz --
more than an order of magnitude
smaller than the 
60-Hz 
final 
measurement
uncertainty. 
This 
final
uncertainty is  
one part in 
$30\,000$
of the natural width of the 
$2^3$P states. 
For such precise measurements,
our standard is to 
continue testing systematic 
effects until the statistical
uncertainty is at least a 
factor of ten smaller than 
the final uncertainty for the measurement.
Doing so allows
for extensive searches
for systematic effects by taking 
data of sufficient precision over
a wide range of experimental parameters.

The measurement was performed blind
by adding an unknown offset 
(of between 
$-4$
and
$+4$~kHz)
to all frequencies
during analysis.
This unknown offset was
implemented more than 
five years ago 
when this work began
and was
revealed
to the authors only 
after completing all
of the analysis for
the measurement
(less than 
48~hours 
before the submission of this 
work for publication).

The weighted average of the 
results shown in 
Fig.~\ref{fig:Parameters}(b)
is
$26\,616\,955\,018(15)_{\rm e}(6)_{\rm Z}(2)_{\rm D}$~Hz,
where the uncertainties come from the 
extrapolations to 
$P$$=$0
(this uncertainty 
is limited by statistical uncertainties),
the 
Zeeman
shift,
and the 
Doppler
shift. 
Adding the uncertainties in 
quadrature would lead to 
an uncertainty of 
16~Hz.
However, 
we take a more conservative 
tack and base our uncertainty
on the proven level of 
consistency demonstrated for 
a wide range of parameters in 
Fig.~\ref{fig:Parameters},
and conservatively assign a
larger uncertainty of 
$\pm$60~Hz 
to our measurement.
Thus, our final measurement result is
\begin{equation}
[E(2^3{\rm P}_0)-E(2^3{\rm P}_1)]/h
=29\,616\,955\,018(60)~{\rm Hz}.
\end{equation}
Combining this measurement with 
our previous measurement
\cite{Kato2018ultahighprecision}
of the 
$J$$=$$1$-to-$J$$=$$2$
interval leads to a determination
of the
$J$$=$$0$-to-$J$$=$$2$
interval
(which is more 
straightforward 
for theory since
it does not involve
the
$2^3$P$_1$
state which 
has a
singlet-state admixture):
\begin{equation}
[E(2^3{\rm P}_0)-E(2^3{\rm P}_2)]/h
=31\,908\,131\,608(65)~{\rm Hz}.
\end{equation}

\begin{figure}[b!]
\includegraphics[width=3 in]
{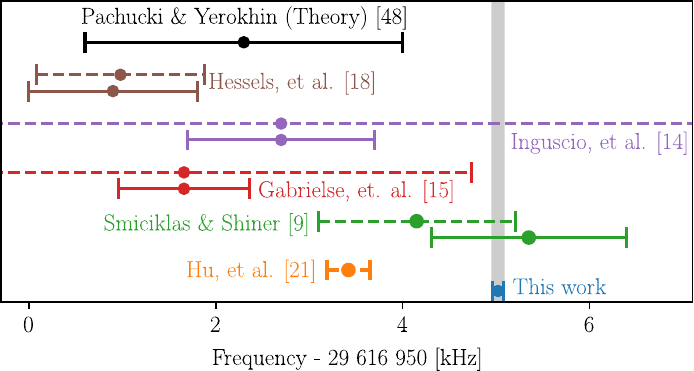}
\caption{\label{fig:Compare} 
A comparison of the present measurement to
previous measurements 
\cite{zheng2017laser,
PRA.91.030502,
PRA.79.060503,
PRL.95.203001,
PRL.105.123001}
and to theory
\cite{PRL.104.070403}.
Corrected 
centers 
including 
quantum-interference effects
\cite{PRA.86.040501,
marsman2015quantum}
are also shown with dashed
error bars.
}
\end{figure}

Our measured value is somewhat 
larger
(1.6~times 
the estimated theoretical uncertainty)
than the best theoretical prediction
\cite{PRL.104.070403},
as seen in 
Fig.~\ref{fig:Compare}.
As can be seen in the figure,
there are large disagreements
with previous measurements.
The measurement of 
Hu, et al.
\cite{zheng2017laser}
disagrees by 
6.7
times their uncertainty.
Our previous 
microwave 
measurement
\cite{PRL.87.173002,PRA.86.012510}
from 
22~years 
ago
also disagrees with the present measurement
(by 4.5 times the uncertainty of 
the previous measurement).
The current work
agrees with the  
measurement of 
Shiner et al.
\cite{PRL.105.123001}.
With the inclusion of quantum interference
corrections
(and the resulting expanded uncertainties)
\cite{marsman2015quantum,
PRA.86.040501},
it is also 
in reasonable agreement with
the 
saturated-absorption 
measurement of 
Gabrielse et al.
\cite{PRL.95.203001}.

Although it is not our 
place to comment on measurements 
made by others, 
we will comment on the 
disagreement with our 
own previous measurement.
That measurement
\cite{PRL.87.173002}
was performed 
without the 
advantages of 
SOF
or
FOSOF
and therefore was
limited to measuring
a single
Lorentzian 
line profile
without the 
possibility of
varying
timing parameters
(as was done here
in 
Fig.~\ref{fig:Parameters}(b-d)).
As a result, 
a systematic effect 
(of undetermined origin)
must have been overlooked
in that measurement.
We note that if we had
been as conservative then
as we have chosen to be now
(by expanding our 
16 Hz 
uncertainty 
to
60 Hz
during our
blind analysis)
there would have been
no discrepancy between our
two measurements.

The current work is the most precise
measurement to date of helium fine structure
and represents a major advance in this precision.
The outstanding 
signal-to-noise
ratio has allowed for a very extensive survey of 
systematic effects.
This work, 
when combined with more precise theory,
could provide 
ppb
tests of the physics and constants 
relevant to the interval --
including a precise determination of the 
fine-structure constant,
the most precise test of 
QED 
in a multi-electron system,
and tests for physics beyond the 
Standard Model.

\section*{Acknowlegements}

This work is supported by 
NSERC, 
CRC, 
ORF, 
CFI, 
NIST 
and a 
York University Research Chair. 

\bibliography{He2018SmallInterval}

\end{document}